\documentstyle[aps,prl,twocolumn,epsfig]{revtex}
\begin{document}
\title{ Multiplicity Moments and Hard Processes in Relativistic Heavy Ion 
        Collisions }
\author{Pengfei Zhuang\\
        Physics Department, Tsinghua University, Beijing 100084, China }
\maketitle

\begin{abstract}
The normalized multiplicity moments and their relation with soft and hard 
processes in relativistic heavy ion collisions are analyzed in a general
two-component model. It is found that, the strong fluctuations in binary 
collision number $N_c$ in minimum-bias events can enhance the hard 
component, especially for  
the higher order moments. This enhancement can not be effectively 
described by
modifying the participant number in the one-component model.    
    
\end{abstract}

\section { Introduction }
The relativistic heavy ion collisions at SPS and RHIC may be the only way to
create the extreme conditions necessary to produce a new state of matter ---
Quark-Gluon Plasma (QGP) in the laboratory\cite{gyu,hei}. 
One can attempt to understand the
energy density achieved in the collisions by studying the multiplicity and
transverse energy distributions through hydrodynamic models\cite{lan}. 
At SPS energies, 
the global quantities like average multiplicity, multiplicity distribution 
and rapidity distribution can be well described by soft processes only, namely
by the number of participant nucleons only\cite{jon}. 
However, at RHIC energies, the
measured pseudorapidity density normalized per participant pair for central
Au-Au collisions shows that $70\%$ more particles are produced than at 
SPS\cite{bac,ste}. This indicates that the yield 
of particles created by hard scattering processes becomes important at
RHIC\cite{gyu,gyul}. 
One can decompose the multiplicity at fixed impact parameter into
a soft component and a hard component as\cite{gyu,ste,kha}
\begin{equation}
\label{softhard}
n = a N_p + b N_c\ ,
\end{equation}
where $N_p$ and $N_c$ are the participant number and binary collision number,
respectively. 

However, the two-component expression (\ref{softhard}) can be effectively 
described by a simple power-law form,
\begin{equation}
\label{soft}
n = c N_p^\alpha\ ,\ \ \alpha > 1\ ,
\end{equation}
which is then similar to that measured at SPS\cite{agg}. 
A natural question then arises:
Can one find other global observables which are more sensitive to the hard 
processes than the multiplicity itself, and which can not be effectively 
described in the models with only soft processes?

As is well known, the multiplicity moments are important characteristics in
multiparticle production. The properties of the multiplicity distribution 
can be completely described by the normalized moments
\begin{equation}
\label{ci}
C_i = {\langle n^i\rangle \over \langle n\rangle ^i}\ ,\ \ i=2,3,\cdots
\end{equation}
In Ref.\cite{zhu} $C_i$ were investigated at SPS energies with a general wounded 
nucleon model. It was found that the normalized multiplicity moments are 
independent of the concrete behavior of elementary nucleon-nucleon collisions,
but dominated by the normalized participant moments
\begin{equation}
\label{cn}
C_i \simeq C_{ip} = {\langle N_p^i\rangle \over \langle N_p\rangle ^i}\ ,
\end{equation}
provided that the colliding nuclei are not too light. 

In this paper we investigate how the hard processes change the normalized 
multiplicity moments. We extend the study in Ref.\cite{zhu} 
to including the hard 
component. We will focus on the sensitivity of $C_i$  
to the colliding energy, nuclear geometry and especially to the geometry 
fluctuations.

\section { Multiplicity Moments }
At fixed impact parameter $b$, the nuclear geometry of soft and hard 
processes is expressed in terms of $N_p$ and $N_c$\cite{khar}, respectively, 
\begin{eqnarray}
\label{npnc}
N_p(b) = \int d^2{\bf s} 
&\Big[&
T_A({\bf s})\left(1-e^{-\sigma_N T_B({\bf b-s})}\right)\nonumber\\
&+&
T_B({\bf b-s})\left(1-e^{-\sigma_N T_A({\bf s})}\right)\Big]\ ,\nonumber\\
N_c(b) = \int d^2{\bf s} 
&&
\sigma_N T_A({\bf s})T_B({\bf b-s})\ ,
\end{eqnarray}
where $\sigma_N$ is the nucleon-nucleon inelastic cross section, and 
$T_A({\bf s})$ and $T_B({\bf b-s})$ are the local participant densities 
in the plane orthogonal to the collision axis defined as
\begin{eqnarray}
\label{tatb}
T_A({\bf s}) &=& \int dz \rho_A ({\bf s},z)\ ,\nonumber\\
T_B({\bf b-s}) &=& \int dz \rho_B ({\bf b-s},z).
\end{eqnarray}

If the average multiplicity distribution of each soft source is 
$g_p(n_p)$, and the average multiplicity distribution of each hard source  
is $g_c(n_c)$, the multiplicity distribution of an $AB$  
collision at impact parameter $b$  
is the supposition of the contributions of $N_p$ soft sources and
$N_c$ hard sources: 
\begin{eqnarray}
\label{gnpnc}
G_{N_p,N_c}(n) 
&=&
\sum_{\begin{array}{c}n_p^{(1)},\cdots, n_p^{(N_p)}\\
                      n_c^{(1)},\cdots,n_c^{(N_c)}\end{array}}
\delta\left(n-\sum_{i=1}^{N_p} n_p^{(i)} -\sum_{j=1}^{N_c} n_c^{(j)}\right)
\nonumber\\
&\times&
\prod_{i=1}^{N_p} g_p(n_p^{(i)})\prod_{j=1}^{N_c}g_c(n_c^{(j)})\ .
\end{eqnarray}
It will be seen later that the results in this paper are not concerned with 
the concrete form of $g_p(n_p)$ and $g_c(n_c)$. Taking into account all the 
processes with different $b$, the final-state multiplicity distribution 
of the nucleus-nucleus collisions is  
\begin{equation}
\label{pn}
P(n) = \sum_{N_p,N_c}p(N_p,N_c)G_{N_p,N_c}(n)\ ,
\end{equation}
where $p(N_p,N_c)$ is the distribution function of $N_p$ and $N_c$. 
Since $N_c$ calculated with (\ref{npnc}) is a monotonous function of $N_p$, 
$p(N_p,N_c(N_p))$ is just the distribution function $p(N_p)$ which can be
obtained from (\ref{npnc}) as
\begin{equation}
\label{pnp}
p(N_p,N_c(N_p)) = p(N_p) \sim b(N_p){db\over dN_p}\ .
\end{equation}

We introduce now generating functions\cite{hov,zhu} $F(\theta), f_p(\theta)$ and 
$f_c(\theta)$ for the whole system and the elementary soft and hard sources,
\begin{eqnarray}
\label{ffpfc}
F(\theta) &\equiv& \sum_n \theta^n P(n)\nonumber\\
          &=& \sum_{N_p}p(N_p)\left[f_p(\theta)\right]^{N_p}
              \left[f_c(\theta)\right]^{N_c}\ ,\nonumber\\
f_p(\theta) &\equiv& \sum_{n_p} \theta^{n_p} g_p(n_p)\ ,\nonumber\\
f_c(\theta) &\equiv& \sum_{n_c} \theta^{n_c} g_c(n_c)\ , \ \ -1\le \theta \le 1\ .
\end{eqnarray}
Differentiating (\ref{ffpfc}) with respect to $\theta$ and making use of the 
relations,
\begin{eqnarray}
\label{average}
&& F(\theta)|_{\theta=1} = f_p(\theta)|_{\theta=1} 
   = f_c(\theta)|_{\theta = 1}=1\ ,\nonumber\\
&& {\partial \over \partial \theta}F(\theta)|_{\theta=1} =\langle n\rangle\ , 
   \nonumber\\
&& {\partial \over \partial \theta}f_p (\theta)|_{\theta=1} 
   =\langle n_p\rangle,\nonumber\\ 
&& {\partial \over \partial \theta}f_c (\theta)|_{\theta=1} 
   =\langle n_c\rangle\ ,\nonumber\\
&& {\partial^2 \over \partial \theta^2}F(\theta)|_{\theta=1} 
   =\langle n(n-1)\rangle\ ,\nonumber\\  
&& {\partial^2 \over \partial \theta^2}f_p(\theta)|_{\theta=1} 
   =\langle n_p(n_p-1)\rangle\ ,\nonumber\\  
&& {\partial^2 \over \partial \theta^2}f_c(\theta)|_{\theta=1} 
   =\langle n_c(n_c-1)\rangle\ ,\cdots
\end{eqnarray}
we derive the multiplicity moments $\langle n^i\rangle$ in terms of the 
elementary soft and hard moments $\langle n_p^i\rangle$ and 
$\langle n_c^i\rangle$ and the nuclear geometry moments 
$\langle N_p^i\rangle, \langle N_c^i\rangle$ and
$\langle N_p^i N_c^j\rangle$,
\begin{eqnarray}
\label{n123}
\langle n \rangle 
&=& 
\langle N_p\rangle \langle n_p\rangle+\langle N_c\rangle \langle n_c\rangle\ ,
\nonumber\\
\langle n^2 \rangle 
&=& 
\left(\langle N_p^2\rangle - \langle N_p\rangle\right) \langle n_p \rangle^2 + 
\langle N_p\rangle\langle n_p^2\rangle\nonumber\\
&+&
\left(\langle N_c^2\rangle - \langle N_c\rangle\right) \langle n_c \rangle^2 + 
\langle N_c\rangle\langle n_c^2\rangle\nonumber\\
&+&
2\langle N_p N_c\rangle\langle n_p\rangle\langle n_c\rangle\ ,\nonumber\\
\langle n^3 \rangle 
&=&
\left(\langle N_p^3\rangle - 3\langle N_p^2\rangle+2\langle N_p\rangle\right) 
\langle n_p \rangle^3\nonumber\\  
&+&
3\left(\langle N_p^2\rangle-\langle N_p\rangle\right)\langle n_p\rangle\langle
n_p^2\rangle +\langle N_p\rangle \langle n_p^3\rangle\nonumber\\ 
&+&
\left(\langle N_c^3\rangle - 3\langle N_c^2\rangle+2\langle N_c\rangle\right) 
\langle n_c \rangle^3\nonumber\\
&+&
3\left(\langle N_c^2\rangle-\langle N_c\rangle\right)\langle n_c\rangle\langle
n_c^2\rangle +\langle N_c\rangle \langle n_c^3\rangle\nonumber\\
&+&
3\left(\langle N_p^2 N_c\rangle - \langle N_p N_c\rangle\right) \langle n_p 
\rangle^2\langle n_c\rangle\nonumber\\
&+&
3\left(\langle N_p N_c^2\rangle - \langle N_p N_c\rangle\right) \langle n_p 
\rangle\langle n_c\rangle^2\nonumber\\
&+&
3\langle N_p N_c\rangle\langle n_p^2\rangle\langle n_c\rangle+3\langle N_p N_c
\rangle\langle n_p\rangle\langle n_c^2 \rangle \ ,\nonumber\\
\cdots\cdots &&
\end{eqnarray}
with the definition of the moments,
\begin{eqnarray}
\label{moment}
&& \langle n^i\rangle = \sum_n n^i P(n)\ ,\nonumber\\
&& \langle n_p^i\rangle = \sum_{n_p} n_p^i g(n_p)\ ,\nonumber\\ 
&& \langle n_c^i\rangle = \sum_{n_c} n_c^i g(n_c)\ ,\nonumber\\
&& \langle N_p^i\rangle = \sum_{N_p\ge N_{min}} N_p^i p(N_p)\ ,\nonumber\\ 
&& \langle N_c^i\rangle = \sum_{N_p\ge N_{min}} N_c^i(N_p) p(N_p)\ ,\nonumber\\
&& \langle N_p^i N_c^j\rangle = \sum_{N_p\ge N_{min}} N_p^i N_c^j(N_p)p(N_p)\ ,
\end{eqnarray}
where we have used the minimum participant number $N_{min}$ to select events. 
$N_{min} = 2$ means minimum-bias events and very large $N_{min}$ corresponds
to central events.

With the known multiplicity moments, the normalized moments 
$C_i=\langle n^i\rangle/\langle n\rangle^i$ 
can be expressed as an expansion in the inverse number of average 
participants $1/\langle N_p\rangle$, 
\begin{equation}
\label{cis}
C_i = {\langle \left({N_p\over \langle N_p\rangle}+{N_c\over 
       \langle N_c\rangle}x
       \right)^i\rangle\over (1+x)^i}
      + {\cal O}\left({1\over \langle N_p\rangle}\right)\ ,
\end{equation} 
where the average ratio of hard to soft component
\begin{equation}
\label{x} 
x={\langle N_c\rangle\langle n_c\rangle\over
\langle N_p\rangle\langle n_p\rangle}
\end{equation}
depends on the elementary 
nucleon-nucleon dynamics and the nuclear geometry. 
If we do not consider peripheral interactions alone, $\langle N_p\rangle,
\langle N_c\rangle >> 1$,
we can then consider only the zeroth order in the expansion (\ref{cis}). 
In this case, only the average ratio of hard to soft component remains, 
the other  
dynamics of elementary soft and hard processes hidden in $\langle n_p^i\rangle$
and $\langle n_c^i\rangle$ with $i > 2$ is washed away by the nuclear 
geometry. 

When the hard contribution can be neglected, namely $x \rightarrow 0$, 
the normalized multiplicity moments are just the normalized 
participant moments, 
\begin{equation}
\label{cci}
C_i = C_{ip} = {\langle N_p^i\rangle\over\langle N_p\rangle^i}\ .
\end{equation}
This is the case discussed in Ref.\cite{zhu} at SPS energies.

\section {Nuclear Geometry and Energy Dependence of Hard Contribution}

let's first determine the soft and hard components $\langle n_p\rangle$ and
$\langle n_c\rangle$ in elementary nucleon-nucleon collisions. To this end,
we compare the average multiplicity with the experimental data for central 
$Au-Au$ collisions. Since we did not introduce rapidity dependence in our 
discussion, we consider only the central rapidity region where the data show
a plateau structure for different centrality bins. By comparing the average 
participant number $\langle N_p\rangle$, the average multiplicity per 
participant pair 
\begin{equation}
\label{data1}
{\langle n\rangle \over 0.5\langle N_p\rangle} = 2\langle n_p\rangle 
\left(1+x\right)\ ,
\end{equation}
and the multiplicity for $P\bar P$
\begin{equation}
\label{data2}
\langle n_{P\bar P}\rangle = 2\langle n_p\rangle +\langle n_c\rangle\ 
\end{equation}
with the experimental data\cite{bac,back} 
in the central rapidity region $|\eta|<1$ at RHIC
and the parametrization of the $P\bar P$ data\cite{abe}
\begin{equation}
\label{pp}
\langle n\rangle_{P\bar P} = 2.5-0.25\ln s+0.023\ln^2 s\ ,
\end{equation}
we can determine at different energies the average ratio $x$ 
and the minimum participant number $N_{min}$ which is 
used to select centrality in calculating geometry moments. 
Using a Wood-Saxon distribution,
\begin{equation}
\label{ws}
\rho_A({\bf r}) = {\rho_0\over 1+e^{r-R_A\over a}}\ ,\ \ 
\int d^3{\bf r}\rho_A({\bf r}) = A\ ,
\end{equation}
\begin{table}[ht]
\hspace{+0cm}
\centerline{\epsfxsize=8cm\epsffile{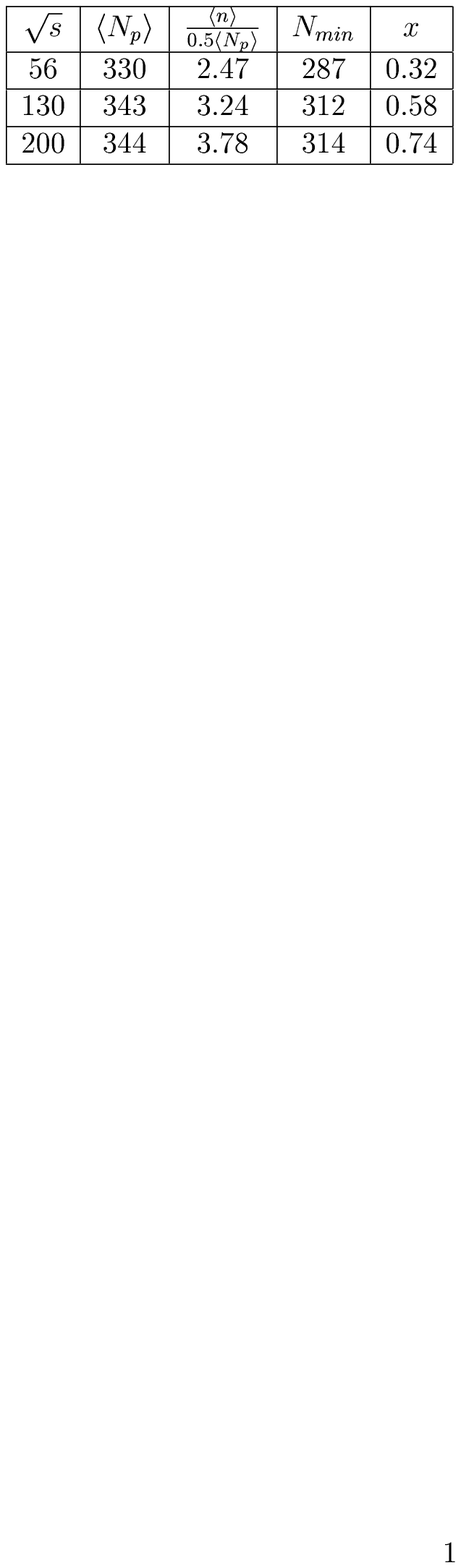}}
\caption{\it The ratio $x$ of hard to soft component and the geometry 
             parameter 
             $N_{min}$ determined from the comparison with the data of central 
             $Au-Au$ collisions at RHIC.}.
\label{tab1}
\end{table}
\noindent with the parameters $a=0.53 fm, \ R_A = 1.1 A^{1/3} fm$ 
for $^{197}Au$ and 
taking $\sigma_N = 37\ mb$ at $\sqrt s = 56 $ A GeV ($\sigma_N = 41\ mb$ for 
$\sqrt s = 130, 200 $ A GeV)\cite{kha}, the two parameters are 
shown in Tab.(\ref{tab1}). We see that at RHIC energies, $x < 1$, the soft 
component is still more important than the hard component. 


The influence of nuclear geometry is twofold: The average numbers 
$\langle N_p\rangle$ and $\langle N_c\rangle$ and the fluctuations of $N_p$ 
and $N_c$ around their average values. For central collisions the average 
numbers $\langle N_p\rangle$ and $\langle N_c\rangle$ are huge, but the 
fluctuations are small. This can be seen clearly in Tab.(\ref{tab1}) where 
$\langle N_p\rangle \ge 330$ and $287\le N_p\le N_p (b=0)$.
For minimum-bias events the
average numbers are relatively small, but the fluctuations are the
maximum. 

\begin{figure}[ht]
\hspace{+0cm}
\centerline{\epsfxsize=8cm\epsffile{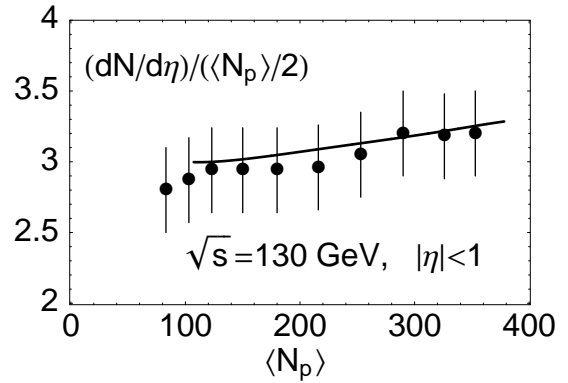}}
\caption{\it The centrality dependence of the average multiplicity 
             normalized to per participant pair and its comparison with
             the RHIC data. }
\label{fig1}
\end{figure}

The multiplicity $\langle n\rangle$ is only related to the average values 
$\langle N_p\rangle$ and $\langle N_c\rangle$. When the 
hard contribution vanishes, the average multiplicity is proportional to
$\langle N_p\rangle$. The hard contribution reflected in the ratio $x$ 
leads to an extra $\langle N_c\rangle$
dependence. The centrality dependence of the average multiplicity per 
participant pair (\ref{data1}) can be calculated by changing the minimum 
participant number $N_{min}$ from $2$ to $N_p (b=0)$. In 
Fig.(\ref{fig1}) it is compared with the data in the central rapidity region 
$|\eta|<1$ for the central $Au-Au$ collisions at $\sqrt s = 130 $ 
A GeV\cite{back1}. The 
extra geometry dependence induced by the hard component is weak.

\begin{figure}[ht]
\hspace{+0cm}
\centerline{\epsfxsize=8cm\epsffile{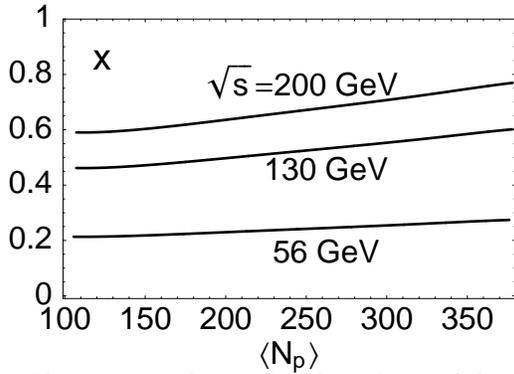}}
\caption{\it The energy and centrality dependence of the average ratio 
             $x$ of hard to soft component.}
\label{fig2}
\end{figure}

Since $\langle N_p^i\rangle = \langle\left(N_p/\langle N_p\rangle
\right)^i\rangle \langle N_p\rangle^i$ and 
$\langle N_c^i\rangle = \langle\left(N_c/\langle N_c\rangle
\right)^i\rangle \langle N_c\rangle^i$, the multiplicity moments 
$\langle n^i\rangle$ for $i\ge 2$ are associated with both the 
average numbers $\langle N_p\rangle$ and $\langle N_c\rangle$ and the 
fluctuations in $N_p$ and $N_c$. From Eq.(\ref{cis}) the normalized moments 
$C_i$ depend on the fluctuations and the average ratio $x$ of hard to soft 
component. Fig.(\ref{fig2}) shows the centrality 
and energy dependence of $x$. At any energy the centrality dependence is 
very weak. Therefore,   
the behavior of the normalized moments $C_i$ is mainly controlled by 
the fluctuations in 
$N_p$ and $N_c$.  Let's first consider the limit of no fluctuations, 
$N_p = \langle N_p\rangle, N_c = \langle N_c\rangle$. In this limit,
\begin{equation}
\label{limit}
 p(N_p) = \delta_{N_p\langle N_p\rangle}\ ,
\end{equation}
we have
\begin{equation}
\label{limit1}
C_i = C_{ip} = 1\ .
\end{equation}
In this case there is no difference between the two-component and one-component
model. Although fluctuations around the average numbers 
always exist, and it is difficult to choose events with the same impact 
parameter $b$, namely with the same $N_p$ and $N_c$, in experiments,  
for very central collisions with large $\langle N_p\rangle$ and
$\langle N_c\rangle$, $N_p$ and $N_c$
fluctuate in a narrow region, the case is then similar to the above limit.

\begin{figure}[ht]
\hspace{+0cm}
\centerline{\epsfxsize=8cm\epsffile{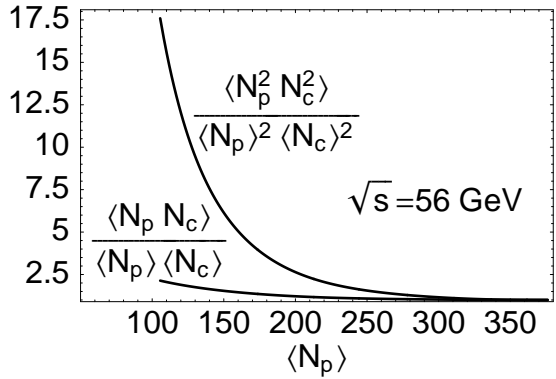}}
\caption{\it The centrality dependence of the geometry fluctuations. }
\label{fig3}
\end{figure}

The fluctuations grow up when the minimum participant number $N_{min}$ 
decreases from its maximum value $N_p(0)$. Fig.(\ref{fig3}) shows the
centrality dependence of the fluctuations $\langle N_p^i N_c^j\rangle/
\langle N_p\rangle^i\langle N_c\rangle^j$. As the orders $i$ and $j$ 
are not too small, the fluctuations are very strong for minimum-bias 
events. 

\begin{figure}[ht]
\hspace{+0cm}
\centerline{\epsfxsize=8cm\epsffile{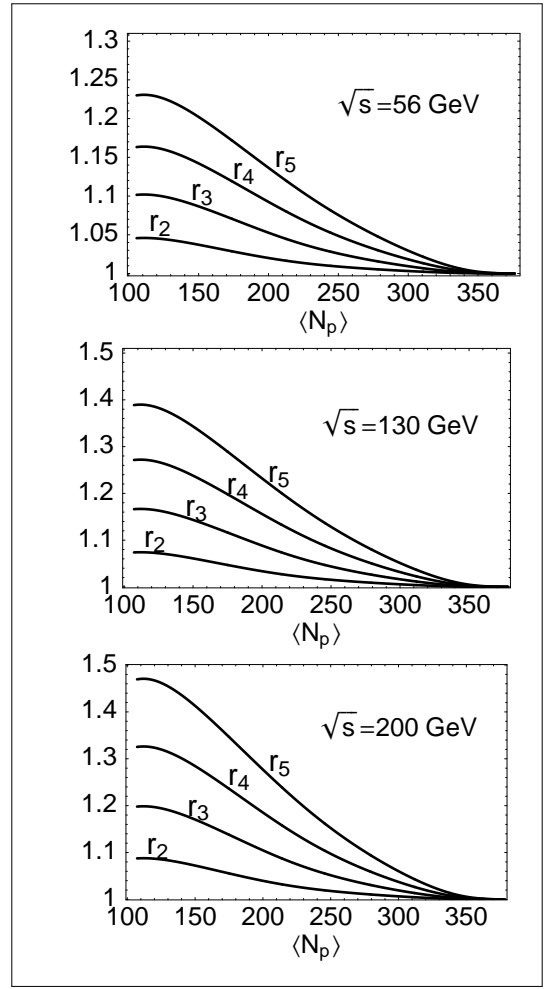}}
\caption{\it The ratio of the two-component to one-component normalized
             moment as a function of the centrality. }
\label{fig4}
\end{figure}

In order to see the contribution from the hard processes, we define the
ratio of the normalized moments with and without consideration of the
hard component,
\begin{equation}
\label{ratioc}
r_i = {C_i\over C_{ip}}\ .
\end{equation}
The centrality and energy dependence of $r_i$ is shown in Fig.(\ref{fig4}).
While there is no remarkable difference between $C_{ip}$ and $C_i$ in central
collisions, the big 
fluctuations in $N_p$ and $N_c$ in minimum-bias events enhance the hard 
contribution, and this enhancement become more and more important when 
colliding energy increases. At $\sqrt s = 200 $A GeV, the hard contribution to 
$C_5$ is almost $50\%$.
 
\section {Comparison with effective model without explicit hard component }
The effect of the hard scattering processes on the average multiplicity can 
be effectively described in the one-component model by modifying the 
participant number\cite{agg},
\begin{equation}
\label{npalpha}
N_c \rightarrow 0\ ,\ \ N_p\rightarrow N_p^\alpha\ ,\ \ \alpha > 1\ .
\end{equation}
By comparing the average multiplicity $\langle n\rangle = \langle N_p^\alpha
\rangle\langle n_p^{eff}\rangle $ with the RHIC data listed in 
Tab.(\ref{tab1}), we can determine the power $\alpha$ and the average 
contribution of each effective soft source $\langle n_p^{eff}\rangle$. 
Corresponding to the colliding energy $\sqrt s = 56, 130, 200$ A GeV, we have
$\alpha = 1.04, 1.07, 1.09 $, respectively. 

In the effective one-component model, the normalized moments are just
the effective participant moments,
\begin{equation}
\label{effci}
C_i^{eff} = {\langle N_p^{\alpha i}\rangle\over \langle N_p^\alpha\rangle^i}\ ,
\end{equation}
\begin{figure}[ht]
\hspace{+0cm}
\centerline{\epsfxsize=8cm\epsffile{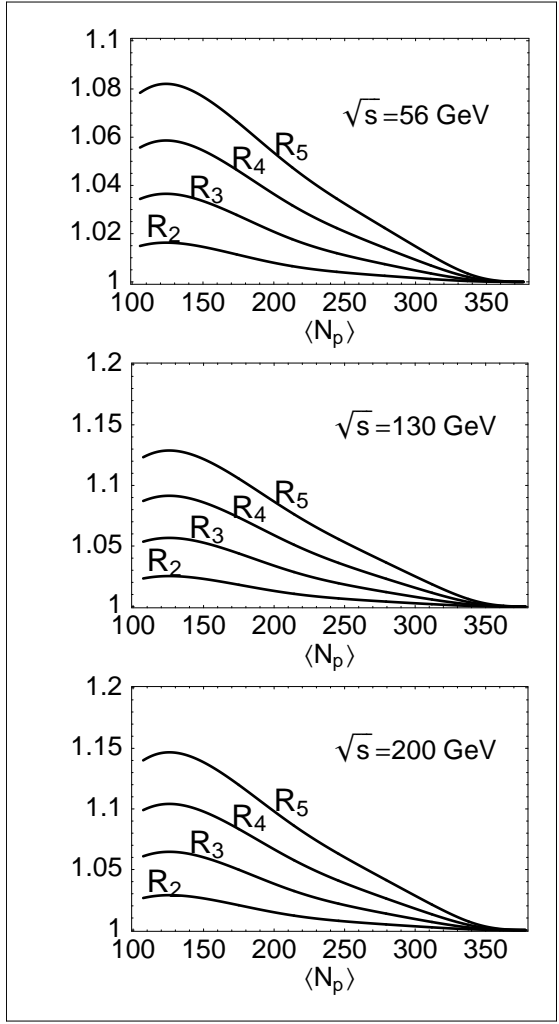}}
\caption{\it The ratio of two-component to effective one-component
             normalized moment as a function of the centrality.}
\label{fig5}
\end{figure}
\noindent when the peripheral interactions are not considered alone. While the 
contribution of the hard processes to the average multiplicity through the
average binary collision number $\langle N_c\rangle$ can be equivalently 
expressed by increasing the average participant number from 
$\langle N_p\rangle$ to $\langle N_p^\alpha\rangle$, the fluctuations in 
$N_c$ can not be effectively included in the fluctuations in 
$\langle N_p^\alpha\rangle$. This can be seen clearly in Fig.(\ref{fig5})
which shows the ratio
\begin{equation}
\label{effratio}
R_i = {C_i\over C_i^{eff}}
\end{equation}
as a function of the centrality for $Au-Au$ collisions. 
From the comparison
with Fig.(\ref{fig4}), $R_i < r_i$, the fluctuations in $N_c$ are partly
included in the fluctuations in the effective participant number 
$N_p^\alpha$. However, the difference 
between the two-component model and the effective one-component model is
still remarkable in minimum-bias events, especially for the higher order 
moments and at high energies. 

\section {Conclusions}
The huge average participant number $\langle N_p\rangle$ and binary collision
number $\langle N_c\rangle$ in relativistic heavy ion collisions make
it difficult to extract dynamic information on hard processes from the 
geometry background. Different from the multiplicity moments 
$\langle n^i\rangle$ which depend on both the average numbers $\langle
N_p\rangle$ and $\langle N_c\rangle$ and the fluctuations in $N_p$ and
$N_c$ strongly, the normalized moments 
$C_i = \langle n^i\rangle/\langle n\rangle^i$ have only weak 
$\langle N_p\rangle$ and $\langle N_c\rangle$ dependence, and are mainly 
associated with the fluctuations in $N_p$ and $N_c$. Therefore, the
geometry background for $C_i$ is not so complicated as that for 
$\langle n^i\rangle$.

We have investigated the normalized moments $C_i$ in the frame of a 
general two-component model. When the hard component can be neglected at SPS
energies, $C_i$ are completely determined by the geometry fluctuations, the
dynamics is totally washed away. When the hard processes become important at
RHIC energies, the average ratio of hard to soft component depends on the
centrality weakly, and $C_i$ are dominated by the fluctuations. For central 
collisions where the fluctuations are weak, $C_i$ approach to $1$, the
dynamic information can not be seen in $C_i$. However, the big fluctuations
in minimum-bias events make us to see clearly the difference between the 
models with and without hard component. 

While the average effect of the hard processes can be effectively described
in the one-component model by modifying the participant number, we have 
found that the fluctuations in the binary collision number can not be
fully included in the fluctuations in the effective participant number.

{\bf acknowledgments}:
The work was supported in part by the NSFC and the Major State Basic Research 
Development Program.

\end{document}